\documentclass[twocolumn,showpacs,preprintnumbers,prl]{revtex4}
\usepackage{amsmath}

\usepackage{graphicx}
\usepackage{dcolumn}
\usepackage{epsf}

\begin{document}

\title{Helical instability of charged vortices in layered superconductors}
\author{A. Gurevich}
\affiliation{National High Magnetic Field Laboratory, Florida State University,
Tallahassee, FL 32310 }
\date{\today }

\begin{abstract}
 It is shown that the electric charge of vortices can result in a helical instability of straight vortex lines in layered superconductors, particularly Bi-based cuprates or organic superconductors. This instability may result in a phase transition to a uniformly twisted vortex state, which could be detected by torque magnetometry, neutron diffraction, electromagnetic or calorimetric measurements.
\end{abstract}

\maketitle

Vortices in superconductors carry the quantized magnetic flux $\phi_0=2\times 10^{-7}$ Oe$\cdot$cm$^{2}$ resulting from the macroscopic phase coherence of superconducting state. Vortices also carry a non-quantized electric charge $q$ caused by the suppression of superconductivity in the vortex core \cite{c1,c2,c3,c4,c5,c6}. In low-$T_c$ s-wave superconductors this charge  is usually negligible and does not manifest itself in the electromagnetic response of vortices driven by the Lorentz force of superconducting currents. However, the situation changes in superconductors with short coherence length $\xi$, low superfluid density and unconventional pairing symmetry combined with the competition of superconductivity with non-superconducting spin or charged ordered states, as characteristic of high-$T_c$ cuprates, recently discovered oxypnictides or organic superconductors \cite{orgsc}. For cuprates, theoretical estimates \cite{c1,c2,c3,c4} predict a relatively large fraction $\sim 10^{-3}$ of the electron charge $e$ per each pancake vortex residing on the $ab$ planes, yet even larger charge of different sign was observed by nuclear quadrupole resonance  \cite{exp}. It has been suggested \cite{c1} that the vortex charge could change the sign of the Hall coefficient observed in cuprates \cite{h} or result in structural transformations of the vortex lattice \cite{halp}.

In this paper we show that vortex charge can cause an intrinsic helical instability of a rectilinear vortex and a phase transition to a twisted vortex state. This instability is different from the helical instability of vortices driven by either currents flowing along the vortex line \cite{spi} or by screw  dislocations \cite{ivlev} or twisted vortex states in rotating liquid He \cite{hel}. The buckling instability of vortices results from the Coulomb repulsion of charged pancake vortices which tend to shift away from the straight line along the c-axis as illustrated by Fig. 1. Such charge fragmentation is inhibited by the vortex line tension caused by weak magnetic and Josephson coupling of vortex pancakes \cite{blat,ehb}, and also by charge screening, which confines the relative  displacements of pancakes on neighboring ab planes within the Thomas-Fermi screening length $\lambda_{TF}$. Thus, the helical instability would be most pronounced in layered materials with low vortex line tension and $\lambda_{TF}\sim \xi$, as characteristic of high-$T_c$ cuprates, ferropnictides or organic superconductors.

To calculate properties of spiral vortices we write the excess linear charge $\rho (r)$ in a vortex as follows
    \begin{equation}
    \rho (r)=\frac{\rho _{0}\xi ^{2}}{r^{2}+\xi ^{2}}+\rho _{a}\exp (-r^{2}/2\xi
    ^{2})  \label{rho}
    \end{equation}
Here the first term is the BCS contribution resulting from
the change in the chemical potential $\mu$ around the core,  $\rho _{BCS}(r)\propto \lbrack \Delta
^{2}(r)-\Delta _{0}^{2}]$, $\Delta (r)\simeq \Delta _{0}r/(r^{2}+\xi ^{2})^{1/2}$ is the modulus of the order
parameter, $\rho _{0}=eN\Delta _{0}^{2}\partial \ln T_{c}/\partial \mu $, $N$ is the density of states at the Fermi
surface in the normal state. The BCS vortex charge $q_{BCS}\simeq 2\pi \xi ^{2}\rho _{0}\ln (\lambda /\xi )$ is spread over
the London penetration depth $\lambda $ \cite{c2}. Strong dependence of the critical temperature $T_{c}$ on doping enhances $\rho _{0}$ in cuprates. The term $\propto \rho_a $ in Eq. (\ref{rho}) is added phenomenologically to take into account the localized core charge due to competing superconducting and antiferromagnetic orders  in unconventional superconductors \cite{c5,c6}. NMR experiments indicate \cite{exp} that the local core charge in cuprates can greatly exceed the BCS contribution.  Eq. (\ref{rho}) corresponds to the following Fourier transform  $\rho (k)=2\pi \xi ^{2}[\rho _{0}K_{0}(k\xi )+\rho _a\exp (-k^{2}\xi^{2}/2)]$ and the total excess charge per unit length $q\simeq 2\pi \xi ^{2}[\rho _{0}\ln (\lambda /\xi )+\rho _a]$ where $K_0(x)$ is the modified Bessel function.
    \begin{figure}                  
    \epsfxsize= 0.7\hsize
    \centerline{
    \vbox{
    \epsffile{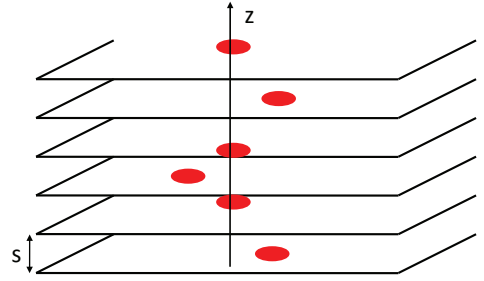}
    }}
    \caption{Spiral instability of a straight chain of charged pancake vortices in a layered superconductor. }
    \label{Fig.1}
    \end{figure}

The excess charge density $\rho (r)$ in a superconductor is
screened in the same way as in a normal metal \cite{c1,c2,c3,c4}. Screening is determined
by the Fourier transform of the static dielectric function $\epsilon (
\mathbf{k}),$ which, for the isotropic Thomas-Fermi model equals $
\epsilon (k)=1+\kappa ^{2}/k^{2}$ where $\kappa=1/\lambda_{TF} $. The Fourier transform of
the electric potential $\varphi(k,p)$ produced by a charged curved vortex parameterized by its
displacement $\mathbf{u}(z)$ relative to the $z-$axis is given by
the Poisson equation:
    \begin{equation}
    (k^{2}+p^{2})\epsilon (k,p)\varphi ({\mathbf{k}},p)=4\pi \rho ({\mathbf{k}}
    )\int_{-\infty }^{\infty }e^{-ipz+i\mathbf{ku}(z)}dz  \label{tf}
    \end{equation}%
where $\epsilon (k,p)$ in a uniaxial material depends on both the in-plane wave vector $k$
and the $z$-component $p$ perpendicular to the layers. From Eq. (\ref{tf}),
we obtain the functional of electrostatic energy $W\{\mathbf{u}
(z)\}=(1/2)\int \rho \varphi d^{3}\mathbf{r}$:
    \begin{equation}
    W=\int \frac{|\rho (k)|^{2}d^{2}\mathbf{k}dpdz_{1}dz_{2}}{4\pi
    ^{2}(k^{2}+p^{2})\epsilon (k,p)}e^{ip(z_{2}-z_{1})+i\mathbf{k}[\mathbf{u}
    (z_{1})-\mathbf{u}(z_{2})]}  \label{W}
    \end{equation}
Here two periodic structures $\mathbf{u}(z)$ are considered:
helical distortions, $u_{x}=u\cos Qz$ and $u_{y}=u\sin Qz$, and planar
zig-zag distortions, $u_{x}=u\cos Qz$ and $u_{y}=0$ where $u$ and $Q$ quantify
the amplitude and the period of the structures. For the spiral
vortex, we have $\mathbf{ku}(z_{1})-\mathbf{ku}(z_{2})=2u\sin
(Qz_{-})[k_{x}\sin (Qz_{+})+k_{y}\cos (Qz_{+})]$ where $z_{\pm }=(z_{1}\pm
z_{2})/2$. Neglecting a possible dependence of $\rho (k,p)$ on $p$ due to
charge modulation along the z-axis, integrating Eq. (\ref{W}) over $z_1+z_2$,
and the polar angle in the $\mathbf{k}$ plane, and adding the elastic energy $F_{e}$,
gives the total line energy of a vortex helix $F_{s}=F_{e}+W_{s}$ where:
    \begin{gather}
    W_{s}=\int_{0}^{\infty }\!\!k|\rho (k)|^{2}dk\!\!\int_{-\infty }^{\infty }\!\!dpdz\frac{%
    e^{-2ipz}J_{0}[2uk|\sin Qz|]}{\pi (k^{2}+p^{2})\epsilon (k,p)},  \label{F}
    \\
    F_{e}=\frac{u^{2}\varepsilon _{0}Q^{2}}{4\gamma ^{2}}\ln \frac{\lambda
    ^{2}\gamma ^{2}}{\xi ^{2}(1+\lambda ^{2}Q^{2})}+
    \frac{\varepsilon _{0}u^{2}}{4\lambda ^{2}}\ln (1+\lambda ^{2}Q^{2})
    \label{fe}
    \end{gather}%
Here $F_{e}$ describes the dispersive tilt energy of a vortex
in a uniaxial superconductor \cite{blat,ehb}, $
J_{0}(x)$ is the Bessel function, $\gamma
=\lambda _{c}/\lambda $ is the anisotropy parameter, and $\varepsilon _{0}=(\phi
_{0}/4\pi \lambda )^{2}$ is the vortex energy scale. For a zig-zag vortex,
we obtain $F_z=F_e/2+W_z$, where
$W_z$ is given by Eq. (\ref{F}) in which $J_{0}[2uk|\sin Qz|]$ is replaced by $J_{0}^2[uk|\sin Qz|]$.
To determine which of the two structures has lower energy, we
minimize $F_{s}$ and $F_{z}$ with respect to $u$ and $Q$ using $\epsilon (k,p)$ for a layered metal \cite{fet}:
    \begin{equation}
    \epsilon (k,p)=\epsilon_0+\frac{\epsilon_0s^{2}\kappa ^{2}\sinh (ks)}{2[\cosh (ks)-\cos (ps)]sk},
    \quad k<2k_{F}  \label{eps}
    \end{equation}%
where $\kappa ^{2}=4\pi^2e^2\hbar^2/m^*\epsilon_0s^2$, $s$
is the interlayer spacing, $\epsilon_0$ is the background dielectric constant,
$m^*$ is the electron effective mass, and $k_{F}$ is the Fermi momentum. For $k>2k_{F}$,
the last factor $k$ in the
denominator should be replaced by $k-\sqrt{k^{2}-4k_{F}^{2}}$. Eq. (\ref{eps})
takes into account the
anisotropy of screening at large $k$, and the Friedel oscillations due to
singularity in $\partial \epsilon /\partial k$ at $k=2k_{F}$. For $(ks,ps)\ll1$,
Eq. (\ref{eps}) reduces to the Thomas-Fermi dielectric function $\epsilon(k,p)=[1+\kappa^2/(k^2+p^2)]\epsilon_0$
with the screening length $\lambda_{TF}=\kappa^{-1}$.
    \begin{figure}                  
    \epsfxsize= 0.7\hsize
    \centerline{
    \vbox{
    \epsffile{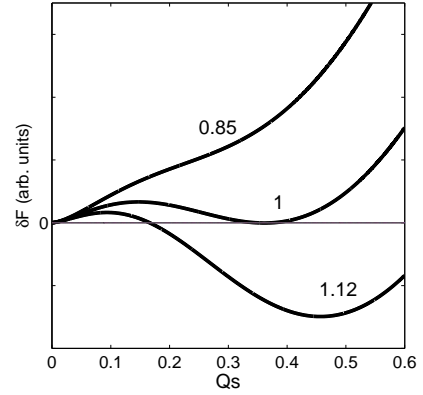}
    }}
    \caption{Energy of the helical vortex line as a function of the wave vector $Q$ and
    different ratios of $q/q_c$ described by eq. (\ref{df}) for $\lambda/s=10^3$ and $(2\lambda_{TF}/s)^2=10$. }
    \label{Fig.2}
    \end{figure}

Now we show that a rectilinear vortex along the c-axis becomes unstable with respect to
bending distortions if $q$ exceeds a critical line charge $q_{c}$. At the instability
threshold $q\approx q_{c}$, Eq. (\ref{F}) can be expanded
in small $u$, and the $z-$ integration produces the $\delta $
functions at $k=0$ and $k=\pm Q$ yielding the following change in $W_e$
    \begin{equation}
    \delta W_{s}=-\frac{u^{2}}{2}\int_{0}^{\infty }\left[ \frac{|\rho (k)|^{2}k}{\epsilon
    (k,0) }-\frac{|\rho (k)|^{2}k^{3}}{(k^{2}+Q^{2})\epsilon (k,Q)}\right]dk
    \label{dW}
    \end{equation}
Hence helical distortions do reduce $W_e$, the electrostatic energy gain increasing as $Q$
increases. The quadratic expansion of Eq. (\ref{W}) for zig-zag distortions yields
$\delta W_{z}=$ $\delta W_{s}/2$. Given that
the charged vortex core is typically larger than either $\lambda_{TF}$ and $s$,
we expand $\epsilon (k,p)$ in $(ks)^{2}\ll 1$ since the integral
in Eq. (\ref{dW}) is mostly determined by the region $k^{2}\ll \kappa ^{2}$, and
$|\rho (k)|^{2}$ rapidly decreases for $k>\xi ^{-1}$.  As the result,
the energy change for small $u$  takes the form
    \begin{equation}
    \delta F_{s}=F_{e}-\frac{q^{2}u^{2}}{4r_{0}^{4}\epsilon_0\kappa ^{2}}\!\!\left[ 1-\frac{
    4\sin ^{2}(Qs/2) }{Q^2s^2+(2Q/\kappa )^{2}\sin ^{2}(Qs/2)}\right]
    \label{df}
    \end{equation}%
Here the effective core radius $r_{0}$ is defined by
$q^{2}/2r_{0}^{4}=\int_{0}^{\infty }k^{3}|\rho (k)|^{2}dk=4\pi
^{2}\int_{0}^{\infty }(\nabla\rho) ^{2}rdr$ and Eq. (\ref{rho}), giving $r_0=\xi$ at $\rho_a\gg\rho_0$.  As $q$ exceeds $q_{c}$,
the function $\delta F_s(Q)$ shown in Fig. 2 first becomes negative at a finite $Q$. Such behavior reflects the
effect of crystalline anisotropy, which strongly reduces the vortex line tension at $Q\lambda \gg 1$ thus facilitating the short wavelength instability. The equation $\partial_Q \delta F_s =0$ at $Q\lambda \gg 1$ yields:
    \begin{equation}
    Q^{2}=\kappa ^{2}/2\ln (\gamma /\xi Q ),  \label{Q}
    \end{equation}%
so that the twist pitch $\ell_s \simeq 2^{3/2}\pi \ln
^{1/2}(\gamma /\xi \kappa )\lambda _{TF}\sim 10\lambda _{TF}$.  For $\lambda_{TF}=0.5-1$ nm
in cuprates \cite{scr},  $\ell_s\simeq 5-10$ nm turns out to be larger than $\xi$.
From the equation $\delta F(q_{c},Q)=0$ and Eqs. (\ref{df})-(\ref{Q}) we obtain the critical charge $q_{c}$
strongly reduced by crystalline anisotropy:
    \begin{equation}
    q_{c}^{2} =2(r_{0}\kappa )^{4}\varepsilon_0\epsilon_0[ \ln
    (\gamma /\xi Q ) +1/2]/\gamma ^{2}.
    \label{qc}
    \end{equation}

Given the relation $\delta F_{s}=2\delta F_{z}$, both helical and zig-zag instabilities occur at the same $q_{c}$ and $Q$, so to see which of these structures has lower energy,
the amplitude of spontaneous distortions $u$ at $q>q_{c}$ is to be calculated. Near the instability threshold $q\approx q_{c}$, the general Eq. (\ref{F}) can be expanded in powers of small $u$ up to
terms $\sim u^{4}$ and integrated at $(r_{0}\kappa )^{2}\gg 1$ as
before. This gives the energy change for the spiral
vortex: $\delta F_{s}/F_{0}=-\alpha _{s}u^{2}+\beta _{s}u^{4}/4$,
where $\alpha _{s}=(1-q_{c}^{2}/q^{2})$, $\beta_{s}=6Q^{2}/r_{0}^{2}(%
\kappa ^{2}+4Q^{2})$, and $F_{0}=q^2Q^2/4r_0^2\kappa^2(\kappa^2+Q^2)\epsilon_0$.
Minimization of $\delta F$ yields the dependence $u(q)$ characteristic of the second order phase
transition:
    \begin{gather}
    u^{2}=\frac{\zeta r_{0}^{2}}{3}\left( 4+\frac{\kappa ^{2}}{Q^{2}}\right) \left( 1-
    \frac{q_{c}^{2}}{q^{2}}\right) \nonumber \\
    \simeq \frac{2\zeta r_{0}^{2}}{3}\left( 2+\ln \frac{\gamma }{\xi \kappa }
    \right) \left( 1-\frac{q_{c}^{2}}{q^{2}}\right).
     \label{u}
    \end{gather}%
where $\zeta=r_0^2\int_0^\infty |\rho(k)|^2k^5dk/\int_0^\infty|\rho(k)|^2k^3dk\to 1$ if $\rho_a\gg\rho_0$. For $q\sim q_c$, the amplitude of the vortex helix is of the order of $\xi$, and the total energy gain equals $\delta
F_{s}=-\alpha _{s}^{2}F_{0}/\beta _{s}$. For a zig-zag vortex, we obtain $\alpha _{z}=\alpha
_{s}/2$ and $\beta _{z}=3\beta _{s}/8$.
Thus, $\delta F_{z}=2\delta F_{s}/3$, so a helical vortex,  which provides the maximum spacing between
charged vortex pancakes at a given $u$,  is more energetically favorable than a zig-zag vortex, which
can lower its energy by transverse buckling distortions.

The instability criterion $q>q_c$ depends on $T$.  For example, the BCS vortex charge $q\propto \Delta_0^2\xi^2$ in Eq. (\ref{rho}) is independent of $T$ at $T_c-T\ll T_c$, while $q_c\propto \varepsilon_0^{1/2}\xi^2\propto (1-T/T_c)^{-1/2}$ in Eq. (\ref{qc}) diverges at $T_c$, suggesting that the helical instability occurs below a certain temperature $T_h<T_c$. However, the NMR experiments \cite{exp} show that the observed $q$ is mostly determined by the non-BCS core contribution modeled by the term $\propto \rho_a$ in Eq. (\ref{rho}). Currently little is known about $\rho_a(T)$, so we analyze the criterion $q>q_c$ at low $T$ where it can be expressed in terms of observable parameters. It is convenient to re-write $q>q_c$ in the form $\eta>\eta_c$ where $\eta_c=qs/e$ is the fraction of the electron charge $e$ per pancake vortex, and
    \begin{equation}
    \eta _{c}=\frac{s(r_{0}\kappa )^{2}\epsilon _{0}^{1/2}}{2^{3/2}\lambda
    \gamma }\left( \frac{\hbar c}{e^{2}}\right) \ln ^{1/2}\left( \frac{
    \gamma}{\kappa \xi }\right) .
    \label{eta}
    \end{equation}
For YBCO with $\epsilon _{0}=25$, $\lambda_{TF}=0.5$ nm \cite{scr}, $\lambda =200$ nm,  $
r_{0}=1.5$ nm, $s=0.85$ nm, and $\gamma=5$, Eq. (\ref{eta}) gives $\eta_c \approx 1.6$, much larger than
$\eta\sim (0.2-2)\times 10^{-2}$ observed for the optimally doped YBa$_2$Cu$_3$O$_7$ \cite{exp}.  Larger values of $\eta\sim (1-5)\times 10^{-2} $ were observed for YBa$_2$Cu$_3$O$_8$ \cite{exp}. The situation becomes more interesting for layered cuprates and organic superconductors, for which $\gamma\sim 100-600$ \cite{orgsc}. For Bi-2212 with $s=1.5$nm, $\lambda=200$nm, $\gamma=500$, $\epsilon_0=10$, and $\kappa r_0=3$, we obtain $\eta_c\simeq 4\times 10^{-2}$ per double $CuO$ planes. Therefore, layered cuprates (particularly underdoped ones) and organic superconductors would be promising candidates for the experimental search for helical vortices, particularly at low $T$ where the vortex core size in the clean limit $r_0(T)\sim \xi(0)T/T_c$ may decrease due to the Kramer-Pesch effect \cite{kramer}. Such core shrinkage strongly reduces $q_c$ in Eq. (\ref{qc}) and could result in an unusual case of  $r_0(T)<\lambda_{TF}$ for which the instability is further enhanced by stronger Coulomb interaction of pancake vortices.

The single vortex helical instability may result in a long-range twist of the interacting vortex lattice. Indeed, if helical displacements $\mathbf{u}(z)$ of all vortices are phase locked, they do not change the flux density, $\nabla\cdot\mathbf{u}=0$ and thus contribute to neither the shear nor the compression energy of the twisted vortex lattice.  Thus, as far as the elastic and electrostatic energies are concerned, vortex structures with a long-range chiral order would be more energetically favorable than structures with different signs of $Q$ or phases of helical distortions on neighboring vortices. In this case the vortex lattice would undergo a phase transition at $q>q(T,B)$ to a uniformly twisted state. Fluctuations and pinning of vortices and proliferation of topological defects may destroy the long range chiral order at higher $T$ and $B$, however if the spacing between pinning centers is much greater than the twist pitch
$\ell$, pinning does not affect the single-vortex helical instability.  The mean-field phase transition at $q=q_c$ results in the specific heat jump $\Delta C=2F_0TB(\partial_T\alpha_s)^2F_0/\phi_0\beta_s=
TB(\kappa^2+4Q^2)(\partial_Tq-\partial_Tq_c)^2/3\phi_0\kappa^2(\kappa^2+Q^2)\epsilon_0$.
If $\partial_T q_c \gg\partial_T q$,  we have $\Delta C/\Delta C_0\sim TB\xi^2\ln(\gamma/\xi\kappa)/T_cB_{c2}\lambda_{TF}^2\gamma^2\epsilon_0$ where $\Delta C_0 = T_c(\partial_T H_c)^2/8\pi$ is the specific heat jump at $T_c$.

Interaction of vortices can be taken into account by adding the elastic twist energy   $B\phi_0u^2Q^2/16\pi(1+Q^2\lambda^2)\simeq \phi_0Bu^2/16\pi\lambda^2$  \cite{blat,ehb} in Eq. (\ref{fe}). Then the problem reduces to the helical instability of a single vortex with a field-dependent line tension ${\tilde \varepsilon_l}=(\varepsilon_0/\gamma^2)\ln(\gamma/\xi Q)+\phi_0 B/8\pi\lambda^2Q^2$ where the last term results from the magnetic cage potential  \cite{blat,ehb}. Minimization of $F(Q)$ at $q=q_c$ and $Qs\lesssim1$ yields
    \begin{eqnarray}
    Q^4\left(2\ln\frac{\gamma}{\xi Q}-1\right)=\left(Q^2+\frac{4\pi B\gamma^2}{\phi_0}\right)\kappa^2,
    \label{Qh} \\
    q_c^2=2\varepsilon_0\epsilon_0 r_0^4\kappa^2(\kappa^2+Q^2)\left(\frac{1}{\gamma^2}\ln\frac{\gamma}{\xi Q}+\frac{2\pi B}{\phi_0Q^2}\right).
    \label{qch}
    \end{eqnarray}
For $B<\phi_0/8\pi\lambda_{TF}^2\gamma^2$, Eqs. (\ref{Qh}) and (\ref{qch}) reduce to Eqs. (\ref{Q}) and (\ref{qc}). For  $B\gg\phi_0/8\pi\lambda_{TF}^2\gamma^2$, we have $Q\sim (B/\phi_0)^{1/4}(\gamma\kappa)^{1/2}$, which gives the critical charge $q_c\simeq(r_0^2/\lambda\lambda_{TF})(\epsilon_0\phi_0B/4\pi)^{1/2}$ independent of anisotropy. The instability region $T<T_h(B)$ defined by $q(T_h)>q_c(T_h,B)$ thus widens as $B$ decreases.
     \begin{figure}                  
    \epsfxsize= 0.63\hsize
    \centerline{
    \vbox{
    \epsffile{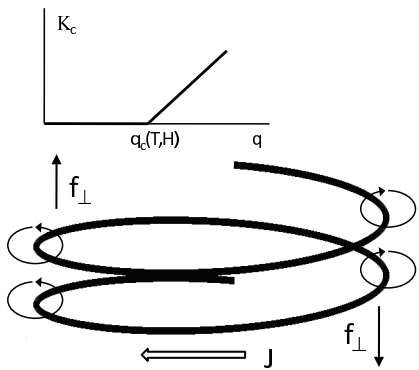}
    }}
    \caption{Mechanism of the torque exerted by a transverse current on the vortex helix where closed lines depict currents circulating around the vortex core. The upper part shows the torque as a function of the line charge
    $q(T,H)$ }
    \label{Fig.3}
    \end{figure}

Helical distortions with $Q\sim \kappa$ can produce minibands in the spectrum of core quasiparticles moving along the vortex. This  may affect the vortex viscosity, vortex mass, and pinning, and also smear the discrete core levels in the STM images of a helical vortex. Vortex chirality also manifests itself in a "fountain"-like currents along the z-axis \cite{ivlev,hel}, and features of flux dynamics controlled by the Lorentz force ${\bf f}=(\phi_0/c)[{\bf J}\times {\bf t}]$ exerted by  the current density ${\bf J}$ per unit vortex length where ${\bf t}(z)=\partial_s{\bf r}/|\partial_s{\bf r}|$ is a tangent unit vector along the vortex helix parameterized by  ${\bf r}=(u\cos Qz, u\sin Qz, z$) and $ds=dz\sqrt{1+Q^2u^2}$. Transport current distorts the helix, yet the net Lorentz force $ \mathbf{F}=(\phi_{0}/c)\int_{0}^{L}[\mathbf{J}\times \mathbf{t}]ds=(\phi _{0}L/c)[\mathbf{J}\times \mathbf{z}]$ is
independent of chirality. The Lorentz forces acting on a helical vortex also produce the torque $\mathbf{\tau }=\int_{0}^{L}[\mathbf{r}\times \mathbf{f}]ds/L$ absent for a straight vortex.
Substituting here ${\bf f}=(\phi_0/c)[{\bf J}\times {\bf t}]$, we obtain that the uniform current density
${\bf J}_\perp$ perpendicular to the helix produces the net torque per unit length along the $z-$axis
    \begin{equation}
    \mathbf{\tau =}\phi _{0}u^{2}[\mathbf{J}_{\bot }
    \times \mathbf{Q}]/2c,  \label{tau}
    \end{equation}
 as illustrated in Fig. 3.  The net torque ${\bf K}={\bf \tau} BV/\phi_0$ exerted by closed magnetization current loops  vanishes, but a uniform current $I$ flowing along a film strip of length $L$ in a perpendicular magnetic field results in the global torque directed along the $y$-axis:
    \begin{equation}
    K_{c}=-u^{2}QBIL /2c.
    \label{ups}
    \end{equation}
Here ${\bf K}_{c}(q)$ exhibits the behavior characteristic of the second order phase transition:
$K_{c}=0$ if $q<q_{c}$ and $K_{c}\propto 1-(q_{c}/q)^{2}$ for $q>q_{c}$ even for the field $H$ directed along the symmetry axis (see Fig. 3). This distinguishes $K_{c}$ from the conventional torque ${\bf K}_a=[{\bf M}\times{\bf H}]$ of tilted straight  vortices in a uniaxial superconductor for which $K_a$ vanishes at ${\bf H}||c$. To estimate the magnitude of $K_{c}$, we compare it with $K_a(\theta)=VH\phi _{0}(1-\gamma
^{-2})\sin 2\theta \ln [\eta H_{c2}/H\varepsilon _{\theta }]/64\pi
^{2}\lambda ^{2}\varepsilon _{\theta }$ for ${\bf H}$ inclined by the angle $\theta$ relative to the c-axis
where $\varepsilon _{\theta }=(\cos^{2}\theta +$ $\gamma ^{-2}\sin ^{2}\theta )^{1/2}$  \cite{vgk}. For $uQ\simeq \kappa \xi >1$, we obtain that $K_c$ exceeds $K_a$ at $J\sim J_{0}\lambda _{TF}/\xi < J_{0}$ for any $\theta$ where  $J_0=c\phi _{0}/16\pi ^{2}\lambda ^{2}\xi $ is of the order of the depairing current density. Thus, the sensitive torque magnetometry could be used to detect twisted vortex structures.

Helical vortices for ${\bf H}$ inclined with respect to the $c$  axis may interfere with the chain and kinked vortex structures in layered superconductors in tilted magnetic fields \cite{aek}. Twisted vortex state may also affect the spiral instability caused by longitudinal currents in the Lorentz force free configurations ${\bf J}||{\bf H}$ \cite{spi}, resulting in asymmetry of the c-axis critical currents parallel and antiparallel to the twist pitch. One could also expect manifestations of the helical overdamped soft modes at $q\simeq q_c$ in the Josephson plasma resonance in layered superconductors at ${\bf H}||c$, and the effect of the chiral mixed state on electrodynamics and the magneto-optical Kerr effect \cite{kerr}.

In conclusion, vortex charge can result in helical vortex instability which can enforce a spontaneous macroscopic twist of the vortex lattice. This can manifest itself in electrodynamic and thermodynamic properties of layered superconductors.

This work was supported by NSF through NSF-DMR-0084173 and by the State of Florida.

\newpage

\end{document}